\documentclass{aa}
\usepackage{graphicx}
\usepackage{natbib}
\usepackage{amssymb}
\bibpunct{(}{)}{;}{a}{}{,} % to follow the A&A style

\begin{document}

\newcommand{\rmd}{{\mathrm d}}
\newcommand{\mm}{\mathrm}
\newcommand{\mi}{\mathit}
\newcommand{\me}{m_\mathrm{e}}
\newcommand{\mpr}{m_\mathrm{p}}
\newcommand{\mk}{\bf}  %markup for changes made
\newcommand{\be}{\begin{equation}}
\newcommand{\ee}{\end{equation}}

\title{X-Ray spectra from accretion disks illuminated by protons}

\author{B.\ Deufel  \and C.P. Dullemond \and H.C.\ Spruit }

\offprints{bed@mpa-garching.mpg.de}
\institute{
       Max-Planck-Institut f\"ur Astrophysik, 
       Karl-Schwarzschildstr.~1, D-85740 Garching, Germany
          }

\date{Received / Accepted }

\abstract{The X-ray spectrum from a cool accretion disk heated by
  virialized protons is computed. The cool disk is either embedded in
  a magnetically heated accretion disk corona or partly extends into
  an ion supported torus (or ADAF).  We calculate the stationary
  equilibrium between proton heating, electron thermal conduction and
  the radiative losses by bremsstrahlung and Compton scattering. A
  heated surface layer on top of the accretion disk is produced with
  temperatures between 60--90 keV above a cool layer with temperatures
  of 0.01 keV (AGN) and 1keV (galactic black hole candidates). The spectra 
  produced by the surface layer are reminiscent of hard state spectra, but a 
  bit too steep, especially for AGN's. Near the inner edge of the disk, where 
the  optical depth of the disk $\tau \lesssim 1$, we find that the cool
  component of the disk disappears.  Instead, the hot protons from the
  corona/ADAF heat the disk, on a dynamical time-scale, to temperatures 
  of several 100 keV, limited by pair production. This region, here called
  a `warm disk',  could contribute significantly to the hard X-ray spectra
  and could be important for feeding material into an ADAF.
   }

\maketitle

\section{Introduction}
\label{sec:intro}

The X-ray spectra of galactic black hole candidates (BH) and active
galactic nuclei (AGN) are remarkably similar, in spite of their large
differences in mass and length scales.  The similarities of the
observed spectra are evident in their power law shape of energy index
$s \approx 1$ in the medium X-ray range, and their high energy cut-off
at $E_c \approx 200$ keV.  It is widely accepted that the power law
spectra are produced by inverse Compton scattering of soft photons on
hot thermal electrons \citep{shapiro76,titar80,pozd83}. Such spectra
depend on the optical depth and the temperature of the Comptonizing
region.  In view of the similarities in the observed power laws these two
parameters apparently vary only within narrow ranges for different sources.

Another common feature evident in many of the observed spectra is the
signature of a fluorescent Fe emission line and a Compton-reflection
component. These two components provide a strong indication that the
X-ray production region is very close to cold matter in the central
regions of BHs or AGNs \citep[e.g.~][]{ross93}.

The classical accretion disk model \citep{sunyaev73} can not explain
the simultaneous presence of cold and hot matter near a compact
object. Another mode of accretion was subsequently proposed by
\cite{shapiro76}. They showed that there is an optically thin hot disk
solution where the plasma is in a two--temperature state, with the
ions near their virial temperature in the inner region of the
accretion disk. This model could in principle account for the
observed Comptonized spectra.  In this model the accretion flow is 
optically thin and radiatively inefficient. It was called an ion supported 
torus by \cite{rees82}. 

The solution described by \cite{shapiro76} turned out to be
thermally unstable. Stable accretion is obtained when the advection
of internal energy with the flow is taken into account \citep{narayan94, 
narayan95a,narayan95b}; these models are called advection
dominated accretion flows (ADAF). Most
of the viscously dissipated energy is advected radially with the flow.
The protons are near their virial temperature, whereas the electrons
are much cooler due to their strong interaction with the radiation
field and the low rate at which they can exchange energy with the
protons via Coulomb interactions. The flow may in principle coexist
with an optically thick, cold accretion disk such that the cold disk
partly extends into the hot torus \citep{esin97}.
 
Another possibility for the simultaneous presence of cold and hot
matter are accretion disk corona (ADC) models
\citep[e.g.][]{nakamura93,haardt91,haardt93,svensson94,dove97}.  
In these models the cold
accretion disk is embedded in a hot corona in a plane--parallel slab
configuration. A large fraction of the gravitational energy is assumed
to be released in the corona by magnetic fields, although the details
of the coronal heating process are still unclear. The protons in the
corona are hot, whereas the disk beneath is cool ($kT\simeq 1$
keV), optically thick, geometrically thin and gas pressure supported.

In both the ADAF and the ADC models there is an energy coupling
between the cold disk material and the hot tenuous plasma above.
Traditionally this interaction was seen in terms of an exchange of
radiation \citep{haardt91,haardt93,hmg93,hmg97}. Here we are interested 
in the energetic coupling via the exchange of matter. We investigate the
penetration of hot protons from the ADC/ADAF into the cool disk, i.e.
`ion illumination'.  The hot protons are stopped in the cold disk 
due to Coulomb interactions with the ambient electrons, rather than
by binary interactions with target ions.

Proton illumination is not restricted to accretion disks.  It was
applied early in X-ray astronomy to explain spectra of galactic X-ray sources 
as neutron star surfaces heated by radially infalling ions
\citep{zel,alme,turolla94}.  Recently \cite{deufel01} revisited the proton
illumination of a neutron star in the context of accretion from an ADAF.

The importance of proton illumination for accretion disks was
suggested by \cite{spruit97} and \cite{spruit00}. Detailed
Monte Carlo calculations of the Comptonization were presented by
\cite{deufel00} [henceforth paper I].  For the present work we have
improved the treatment of the radiative processes by solving the
radiative transfer equation including thermal emission due to
bremsstrahlung. Pair production is also taken into account.  We
further allow for energy redistribution due to thermal {\mk electron} conduction
within the disk.  The interaction between the impinging protons and
the accretion disk is computed time dependently in a one-dimensional,
plane-parallel approximation.  The density distribution is found from
hydrostatic equilibrium including the force exerted by the
decelerating protons.

The cool disk acts as an effective thermalizer for energetic
(Comptonized) photons as long as its optical depth is large enough. At
lower optical depth, such as occurs near the inner edge of the disk,
photon production is less efficient.  Therefore we expect different
solutions for disks of low optical depth. This is borne out by the
results reported in Section \ref{sec:slim}.

\section{A cool disk extending inside an ion supported flow}
\label{sec:hydro2}

\subsection{Optically thick case}

Most of the physics of the interaction between the hot supported flow and
the underlying cool disk is rather independent of the structure of the cool
disk, since it acts mainly as a thermalizing surface. The height above the 
%Kees: zo beter?
midplane where the hot ion-supported flow and the cold disk interact is 
important, however, since this determines the gravitational acceleration, and 
hence the density within the interaction layer. Processes such as photon 
production by bremsstrahlung depend on the electron density explicitly. 
To obtain an estimate for the thickness of the cool disk, the
amount of energy released in the cool disk must be specified.  For this
purpose we adopt an $\alpha$-disk prescription as described in Svensson and
Zdziarski (1994). In this model, a fixed fraction $f$ of the gravitational
energy release takes place above the disk, and a fraction $1-f$ in the cool
disk. For details we refer also to paper I.

In order to simplify our treatment we assume that the vertical disk
structure can be approximated at a given radius by a plane--parallel,
one--dimensional geometry, so that the vertical $z$--coordinate is the
only dimension relevant to our problem. We find the vertical
density distribution from the equation of hydrostatic equilibrium,
\begin{equation}
  \frac{dP}{d\tau}=\frac{\Omega^2_\mm{K} z}{\kappa} + 
  \frac{\partial P_\mm{p}}{\partial\tau}(\tau),
  \label{hydro}
\end{equation}
where $\Omega_\mm{K}=\left(G M/R^3\right)^{1/2}$ is the local
Kepler angular velocity, $\kappa = \kappa_\mm{es}=0.40$ cm$^2$g$^{-1}$ 
is the electron scattering opacity and $\tau$ is the electron scattering
optical depth. Note that we use $\tau$ here as a coordinate, replacing
the vertical height $z$. The first term on the right
represents the usual weight of the disk; the second term takes into
account the momentum exerted by the deceleration of the incident
protons as a function of optical depth.  The force $\partial P_\mm{p}/ 
\partial\tau$ is evaluated by recording the change of velocity $\Delta
v_\mm{p}$ of the incoming protons as a function of optical depth (see
Sect. \ref{sec:prot}). The contribution of the radiative pressure is
neglected as we study cases well below the Eddington luminosity.

Together with the equation for the vertical coordinate $z$,
\begin{equation}
  \frac{dz}{d\tau} = \frac{-k}{\mu\,m_\mm{p}\,\kappa}\frac{T(\tau)}{P(\tau)},
\label{dz}
\end{equation}
we integrate Eq. (\ref{hydro}) via a fourth order Runge-Kutta
method.  We set $\mu=\frac{1}{2}$ for the idealized case of an ionized
hydrogen atmosphere.

The initial value for the pressure at the top of the layer is set to a
small fraction of the coronal pressure, $P_\mm{top}(\tau=0) = 10^{-2}
n_\mm{p}\,k\,T_\mm{vir}$. The geometrical height $z_\mm{top}$ of this
upper boundary layer is not known in advance. We solve the pressure
profile by starting with an initial guess for the height of the disk
above the mid-plane, $z_\mm{top}$.  We integrate Eq. (\ref{hydro}) to
the maximum optical depth of our model atmosphere $\tau_\mm{bot}$ at
the vertical height $z_\mm{bot}$. Usually we set $\tau_\mm{bot}=3$.
At this Thomson optical depth the protons have already lost more than
0.99 of their energy.  At $\tau_\mm{bot}$ pressure balance between the
simulated layer and the underlying cool disk requires
$P(\tau_\mm{bot})=P(z_\mm{bot})$.  We iterate until we find the right
value for which this pressure condition at $z_\mm{bot}$ is fulfilled.
Thus we have matched the simulated layer to the underlying cool disk.
The temperature of the cool disk and the pressure at the mid--plane
are determined as in paper I.

\subsection{Moderate optical depth}

Near the inner edge of a cool Shakura--Sunyaev disk at the radial
distance $R_\mm{i}$ from the compact object, the surface density
drops to low values. In the limit $H/R \ll 1$ (where $H$ is the disk
pressure scale height), the surface density
varies as $[1-(R_{\rm i}/R)^{1/2}]$. Close to the inner edge
of an accretion disk, the assumption of a large optical depth is therefore not
valid any more. A certain minimum optical depth is necessary, however,
for efficient cooling of the disk by soft photons to be possible. In section
\ref{sec:slim} we explore cases where the total optical depth of the layer is
of order unity or less. The procedure to calculate the hydrostatic balance 
in such cases is almost the same as in the cases of a surface layer on top 
of an optically thick disk. But instead of matching the computed layer to 
an underlying cool disk, we put
the bottom of the simulated layer is at $z(\tau_\mm{bot})=0$, the mid-plane 
of the disk. We are interested in simulating the complete disk, i.e. from the 
upper surface to the lower surface.  We do this by introducing an artificial
`mirror' placed at the mid-plane. All relevant physical quantities
of the model are reflected at this mirror.  Thus we have to compute
one half space only instead of simulating the whole vertical disk
structure.

\section{Heating and cooling processes of the model}

\subsection{Proton illumination of the accretion disk}
\label{sec:prot}
\subsubsection{Coulomb interactions in an ionized plasma}
 
At the typical energies of the protons incident on the cool disk, the energy 
loss is mostly by long-range Coulomb interactions with the electrons in the 
disk (small-angle scattering on the large number of electrons in a Debye 
sphere). This is opposite to the case of protons with a temperature near 
that of the plasma in which they move. In the latter case, the equilibration 
among the protons is faster than between electrons and protons, by a  
factor of order $(m_\mm{p}/m_\mm{e})^{1/2}$. To see how this apparent 
contradiction is resolved, consider the basic result for the energy loss of a 
charged particle moving in a fully ionized, charge-neutral plasma. This was 
derived by Spitzer (1962) (making use of Chandrasekhar's (1942)\nocite{chandra} 
earlier results on dynamical friction). Introduce as a measure of distance in the 
plasma the Thomson optical depth $\tau$, i.e. $\rmd \tau=\sigma_{\mm T}n 
\rmd l$, where $\sigma_{\mm T}=8\pi e^4/(3m_\mm{e}^2c^4)$ is the 
Thomson cross section, $n$ the electron density and $l$ the distance. The 
rate of change of energy $E=\mpr v^2/2$ of a proton moving with velocity 
$v$ in a field of particles with charge $e$ and mass $m_\mm{f}$ (the `field 
particles' in Spitzer's nomenclature) is then given by Spitzer's Eq. 5-15. In 
our notation, this can be written in terms of  the energy loss length 
$\tau_\mm{f}$ for interaction with the field particles $\mm f$,

\be 
\tau_\mm{f}^{-1}={1\over E}({\rmd E\over \rmd\tau})_\mm{f}=
3\ln\Lambda ({\me\over\mpr})^2({c\over v})^4(1+{\mpr\over 
m_\mm{f}})F(x_\mm{f}),\label{spit}
\ee
where $\ln\Lambda$ is the Coulomb logarithm (which is determined by 
the size of the Debye sphere). Here
\be F(x)=\psi(x)-x\psi^\prime (x), \ee
where $\psi$ is the error function and $\psi^\prime$ its derivative, and 
$ x_\mm{f}= v[{m_\mm f}/(2kT)]^{1/2}$ is (up to a numerical factor) 
the ratio of the incident proton's velocity to the thermal velocity of the field 
particles. The limiting forms of $F$ are
\be 
F(x)\rightarrow x^3/3\quad (x\rightarrow 0),\qquad F\rightarrow 1\quad 
(x\rightarrow\infty).
\ee
We can evaluate (\ref{spit}) under the assumption that the field particles 
that are most relevant for the energy loss are the protons or the electrons, 
respectively, and compare the loss lengths. If the incident proton has velocity 
comparable with the thermal velocity of the field protons, we have $x_\mm{p} 
\sim 1$, and $x_\mm{e}=(\me/\mpr)^{1/2}x_\mm{p}\ll 1$. $F(x_\mm{p})$ is 
then of order unity and $F(x_\mm{e})\approx x^3_\mm{e}/3$. Setting 
$\mm{f}=\mm{e}$ respectively $\mm{f=p}$ in (\ref{spit}) and taking the ratio, we have
\be {\tau_\mm{e}\over\tau_\mm{p}}\approx ({\mpr\over\me})^{1/2}. \ee
The loss length for interaction with the electrons is thus much longer than for 
interaction with the protons, and the interaction with electrons can be neglected. 
This well known result is the relevant limit for the relaxation of a  proton 
distribution in a plasma that is not too far from its thermal equilibrium.

For incoming protons of high energy, however, the result is different because 
$x_\mm{e}$ is not sufficiently small any more. In the high-$v$ limit, $F(x_\mm{e})=
F(x_\mm{p})=1$, and one has
\be {\tau_\mm{e}\over\tau_\mm{p}}\approx 2{\me\over\mpr}=10^{-3}. \ee
In this limit, the energy loss is thus predominantly to the electrons. A related 
case is that of the ionization losses of fast particles in neutral matter (for 
references see Ryter et al. 1970). The case of an ionized plasma is simpler, 
since the electrons are not bound in atoms. The change from proton-dominated 
loss to electron-dominated loss takes place at an intermediate velocity $v_{\mm c}$, 
at which $F(x_\mm{p})\approx 1$ but $x_\mm{e}$ still small, so that 
$F(x_\mm{e})\approx x^3_\mm{e}/3$. Equating $\tau_\mm{e}$ and 
$\tau_\mm{p}$ then yields
\be x_{\mm e,c}\approx ({\me\over\mpr})^{1/3},\ee
or
\be {E_{\mm c}\over kT}\approx ({\mpr\over\me})^{1/3}\approx 12.\ee
For the electron temperatures we encounter in our models, $T\sim 100$ keV, 
energy loss to the field protons can thus be neglected for incoming protons 
with energy $E\ga 1$ MeV. This is the case in all calculations presented here.

\subsubsection{Corrections at high and low energies}

Spitzer's treatment is non-relativistic, while virialized ADAF protons near 
the hole can reach sub-relativistic temperatures. A fully relativistic treatment 
of the Coulomb interactions in a plasma has been given by Stepney and Guilbert (1983). 
We have compared the classical treatment according to Spitzer's theory with 
this relativistic result in Deufel et al. (2001), and found it to be accurate to better 
than 5\% for proton temperatures $<100$ MeV. The classical approximation 
in Spitzer's analysis therefore does not introduce a significant error for the 
problem considered here. 

For high energies, the Coulomb energy loss becomes so small that loss by 
direct nuclear collisions becomes competitive. This happens (cf. Stepney and 
Guilbert 1983) at $E\ga 300$ MeV, an energy that is not reached by virialized 
protons except in the tail of their distribution. We ignore these direct nuclear 
collisions. Note, however, that a gradual nuclear processing by such collisions 
can be important (Aharonian and Sunyaev, 1984)\nocite{aha}, in particular for the 
production of the Lithium. The Lithium overabundances seen in the companions 
of LMXB (Mart\'{\i}n at al., 1994a)\nocite{martin94a}, may in fact be a 
characteristic signature of the interaction of an ADAF and a disk described here 
(Mart\'{\i}n et al. 1994b, Spruit 1997)\nocite{martin94b}\nocite{spruit97}.

As the protons slow down, they eventually equilibrate with the field protons. 
This last part of the process is not accurately described by the energy loss 
formula (\ref{spit}). In addition to the simple energy loss of a particle moving 
on a straight path through the plasma, one has to take into account the random 
drift in direction and energy resulting from the interaction with the fluctuating 
electric field in the plasma. This drift can be ignored to first order (end of section 
5.2 in Spitzer 1962), but takes over in the final process of equilibration with the 
plasma. This last phase involves negligible energy transfer compared with the 
initial energy of the protons in the present calculations, and can be ignored here.

\subsubsection{Charge balance}

The protons penetrating into the disk imply a current that has to be balanced 
by a `return current'. As in all such situations, this return current results from 
the electric field that builds up due to the proton current. This field
drives a flow of electrons from the ADAF to the disk which maintains the charge
balance. Since the electron density in the disk is high, the return current does
not involve a high field strength.

\subsubsection{Calculations}
The numerical method used to calculate the electron heating rates of the accretion 
disk due to proton illumination is described in detail in paper I;
here we give a short overview. 

We follow the evolution of an initially 
Maxwellian distribution of protons, which we place above the cool accretion 
disk, as the protons penetrate through the disk atmosphere. For
the temperature of the incident protons in our model we take the local virial
temperature 
\begin{equation}
\label{tvir1}
T_\mm{p}=T_\mm{vir}=\frac{G M m_\mm{p}}{3 k_{\mm B} R}\approx
\frac{156}{r}\hspace{2mm}\mbox{MeV},
\end{equation}
where $G$ is the gravitational constant, $M$ the mass of the black
hole, $m_\mm{p}$ the proton mass, $R$ the radius from the central
object and $k_\mm{B}$ is the Boltzmann constant.  $r$ is the
dimensionless radius, $r=R/R_\mm{S}$, where $R_\mm{S} = 2GM/c^2$ is
the Schwarzschild radius.

The energy loss of the incoming protons is computed from (\ref{spit}), with
$\mm f=e$. Denote by $\tau$ the Thomson optical depth 
measured vertically into the layer, from its top.  The rate of change of 
energy with depth for a proton at depth $\tau$ moving at an angle $\theta$ 
with respect to the vertical is then
\be 
{1\over E}{\rmd E\over \rmd\tau}={3\ln\Lambda\over\cos\theta} 
{\me\over\mpr}({c\over v})^4[\psi(x)-x\psi^\prime (x)], \label{eq:dedtau}
\ee
with 
\be x=x_\mm{e}=v({m_\mm e\over 2kT})^{1/2},\ee
while $T(\tau)$ is the depth-dependent temperature of the layer, {\mk and
$v$ the depth and angle-dependent velocity of the incident proton. The
angular distribution of the protons is discretized in 50 equidistant points}.
The Debye-length in the plasma, measured in Thomson optical depths, is 
quite small so that temperature variations over a Debye-length can be 
ignored and the local energy loss rate at any depth is adequately described
by (\ref{eq:dedtau}).

The energy flux of the incident flux $q_{\rm p}$ is proportional to
the density in the ADAF, which depends on its accretion rate and
viscosity. Instead of using detailed models we parameterize these
dependences by scaling the energy flux of the incident protons with
the local energy dissipation rate in the ADAF. Using the thin-disk
expression for this dissipation rate, we set
\begin{equation}
\label{eflux}
q_\mm{p}=fQ(R) = f\cdot\frac{3GM\dot{M}}{8\pi R^3}\cdot J(R),
\end{equation}
where $f$ is an adjustable numerical factor of order unity and
$J(R)=1-(R_\mm{i}/R)^{1/2}$.
This scaling of the proton flux is also useful if one has a
coronal model in mind instead of an ADAF. The factor $f$ then is
interpreted as the fraction of the viscous energy release that is
dissipated in the corona. 

We follow the protons from the corona numerically through the disk
atmosphere and record the energy loss as a function of optical depth
$\tau$ according to Eq. (\ref{eq:dedtau}). 
This yields the local time dependent heating rate
$\Lambda_\mm{P}(\tau)$.

\subsection{Radiative transfer}
\label{sec:radiative}

We treat the radiative transfer by solving the radiative transfer
equation
\begin{equation}\label{eq-transfer-eq}
\mu\frac{dI_{\mu,\nu}}{dz} = j^\mi{ff}_{\nu} + j^{c}_{\mu,\nu} 
- (\alpha^\mi{ff}_{\nu}+\alpha^{c}_{\nu}) I_{\mu,\nu}
\end{equation}
where $I_{\mu,\nu}$ is the intensity as a function of frequency $\nu$
and photon angle $\mu=\cos(\theta)$, $\alpha^\mi{ff}_{\nu}$ is the
bremsstrahlung absorption coefficient, $\alpha^{c}_{\nu}$  the
Compton extinction coefficient, $j^\mi{ff}_{\nu}$ the
bremsstrahlung emissivity and $j^{c}_{\mu,\nu}$ the Compton
emissivity. The method for solving the radiative transfer equation is
described in \cite{deufel01}. From the solution of the radiative
transfer equation we obtain the radiative cooling rates
$\Lambda_\mm{rad}(\tau)$.

\subsection{Pair processes}

As discussed below, the cases of moderate optical depth can become
rather hot so that pair processes have to be included.  We include
photon--photon pair production ($\gamma\gamma\rightarrow e^{+}e^{-}$)
as well as pair production due to $ee$ collisions ($ee\rightarrow e e
e^{+} e^{-}$) in a steady state, i.e. the pair production rate equals
the pair annihilation rate.

The photon-photon pair production rate can be expressed as a quadruple
integral over the dimensionless photon energies $x_1=h\nu_1/m_ec^2$,
$x_2=h\nu_2/m_ec^2$ and photon angles $\mu_1$, $\mu_2$, respectively
\citep[cf.~][]{zane95}:

\begin{equation}
  R_{\gamma\gamma} = \frac{4\pi}{h^2c} \int 
dx_1 dx_2 d\mu_1 d\mu_2 \frac{I_{\mu_1,\nu_1}I_{\mu_2,\nu_2}}{x_1^2x_2^2}
F(x_{+},x_{-}),
\end{equation}
where $F(x_{+},x_{-})$ is defined as

\begin{equation}
F(x_{+},x_{-}) = \int_{x_{-}}^{x_{+}}
\frac{\sigma(x)\,x^3dx}{\sqrt{(x_{+}^2-x^2)(x^2-x_{-}^2)}},
\end{equation}
with $x_{\pm}^2\equiv x_1x_2[1-\cos(\theta_1\pm\theta_2)]/2$
\citep{stepg83}. The pair cross-section $\sigma(x)$ is \citep{jauch76}

\begin{equation}
\sigma(x) = \frac{1}{2}(1-\beta^2)\left[(3-\beta^4)\log\left(
\frac{1+\beta}{1-\beta}\right)-2\beta(2-\beta^2)\right],
\end{equation}
with $\beta\equiv\sqrt{x^2-1}/x$.

In our models we have simplified the above integral by replacing the
intensities by their corresponding mean-intensities:
$I_{\mu_1,\nu_1}\rightarrow J_{\nu_1}$ and $I_{\mu_2,\nu_2}\rightarrow
J_{\nu_2}$. For any given value of $y\equiv x_1x_2$ we can now
evaluate the function $G(y)\equiv\int F(x_{+},x_{-}) d\mu_1 d\mu_2$,
which is independent of the radiation field, and which can be
tabulated and stored beforehand. The pair production rate now reduces
to
\begin{equation}
R_{\gamma\gamma} \approx \int G(x_1x_2) J_{\nu_1}J_{\nu_2} dx_1 dx_2\;.
\end{equation}
Thus the quadruple integral is reduced to a double integral, which is much
faster to evaluate. An accurate evaluation of this integral can be made by
writing $dx_1dx_2=x_1x_2 d\log(\sqrt{x_1x_2}) d\log(x_1/x_2)$, since the
function $G(x_1x_2)$ is independent of $x_1/x_2$.
 
The pair production rate due to electron--electron collisions is
e.g. given in \cite{svensson82} as

\begin{equation}
  \dot{n}_{+}^{ee}=(n_{+}+n_{-})^2 c \,r_e^2 \,P_{ee}(\theta)\,
\label{eq:ee}
\end{equation}
where we have introduced $\theta = k T_\mm{e}/ m_\mm{e}c^2$ and

\begin{equation}
  \label{eq:pee}
  P_{ee} = \frac{1}{2}\,\alpha_f^2\,\frac{28}{27\pi}(2 \ln 2 \theta)^3
\end{equation}
Here $\alpha_f$ is the fine structure constant. This production rate is
only included for $\theta>0.6$.

The pair annihilation
rate is also given by \cite{svensson82} as

\begin{equation}
  \label{eq:ann}
  \dot{n}_{+}^\mm{an} = n_{+}\,n_{-}\,c\,r_e^2\,A(\theta)\,,
\end{equation}
where

\begin{equation}
  \label{eq:at}
  A(\theta)=\frac{\pi}{1+2\theta^2/\ln(2\eta\theta + 1.3)}\,,
\end{equation}
$\eta\simeq 0.56146$. Now we can set up the pair balance equation in
the stationary case,

\begin{equation}
  \label{eq:pbal}
  R_{\gamma\gamma} + n^2\,c\,r_e^2 \; [ (2z+1)^2  P(\theta) -
  z(1+z)  A(\theta)] = 0.
\end{equation}
We have used the charge neutrality condition 
$n_{-} =  n_{+} +\,n$ and $z = n_{+}/n$ is the ratio of the pair number
density to the proton number density $n$. Eq. (\ref{eq:pbal}) is a
quadratic equation for the pair fraction $z$. The positive root of
this equation is

\begin{equation}
  \label{eq:root}
  z(\tau)=\frac{1}{2}\left[-1 + \frac{\sqrt{A(\theta)\,c\,n^2\,r_e^2 +
        4R_{\gamma\gamma} } }
    {n\,r_e\,\sqrt{c}\sqrt{A(\theta) - 4P(\theta)}}    \right] 
\end{equation}

The additional electrons from pair production serve as further
scattering partners for the Coulomb collisions and the Compton
scattering.  Effects due to electron--positron bremsstrahlung are not
included nor is a treatment of the annihilation line.

\subsection{Energy balance from heating and cooling}

We begin our computation with an isothermal temperature profile in
hydrostatic equilibrium according to Sect. \ref{sec:hydro2}.  The
initial temperature throughout the layer is $T_\mm{d}$ (see paper I)
for the thick disks and we set $T_\mm{e}=1$ keV for disks with
moderate optical depths.  The solutions do not depend on the initial
temperature profile. First we calculate the heating rates
$\Lambda_\mm{p}^+(\tau)$ from the Coulomb interactions and the cooling
rates $\Lambda_\mm{rad}^-(\tau)$ due to the radiative processes
bremsstrahlung and multiple Compton scattering. Additionally the
energy redistribution due to electron thermal conductivity is included
{\mk using Spitzer's classical value, as in \cite{deufel01}. This process
adds the contribution $\Lambda_\mm{cond}(\tau)$ to the energy balance. 
The validity of classical electron conduction was checked by evaluating the
electron mean free path $\lambda$, which turns out to be of the order
$10^2$ cm or less in the black hole candidate cases shown in Fig.\ \ref{fig:coolall}. 
This is much smaller than the temperature gradient length (more than $10^4$ cm),
so the condition for validity of the electron formula used is amply satisfied.
The same holds for the AGN cases.}

The equilibrium is computed following the time evolution of the layer
until a balance between the heating and cooling processes is obtained.
In the optically thick models, the time scales for approaching
equilibrium turn out to be a sharp function of depth in the model. To
deal with this, an adaptive time stepping process is used in which the
time step depends on both time and depth in the model. Stability of
this process was obtained by scaling this step with the shortest of
the energy exchange time scales associated with the contributing
heating and cooling processes. With this procedure the time evolution
of the model is not realistic, but the final equilibrium obtained is.
For the cases with moderate optical depth we use a depth-independent
time step. In these cases, the time evolution of the model is also
realistic.

The total change of enthalpy per time step as function of optical
depth can then be expressed by

\begin{equation}
  \label{ent} \frac{\Delta w(\tau)}{\Delta t} = \rho c_p \frac{\Delta
T(\tau)}{\Delta t} = \Lambda_\mm{p}^+(\tau) + \Lambda_\mm{rad}^-(\tau)
+ \Lambda_\mm{cond}(\tau)\;,
\end{equation}
where $c_p$ is the specific heat at constant pressure.

Now the change of temperature as a function of optical depth can be
calculated. With the new temperature profile the hydrostatic structure
is updated according to Sect.  \ref{sec:hydro2}.  We follow the
simulation until the Coulomb heating is balanced by the radiative
cooling and energy redistribution due to electron conductivity. The
energy balance of all our computations is better than 10\%.

\section{Results of the model computations}

\subsection{Proton illumination of cool disks in BH and AGN}
\label{sec:protill}

\begin{figure*}[th]
  \sidecaption \includegraphics[width=12cm]{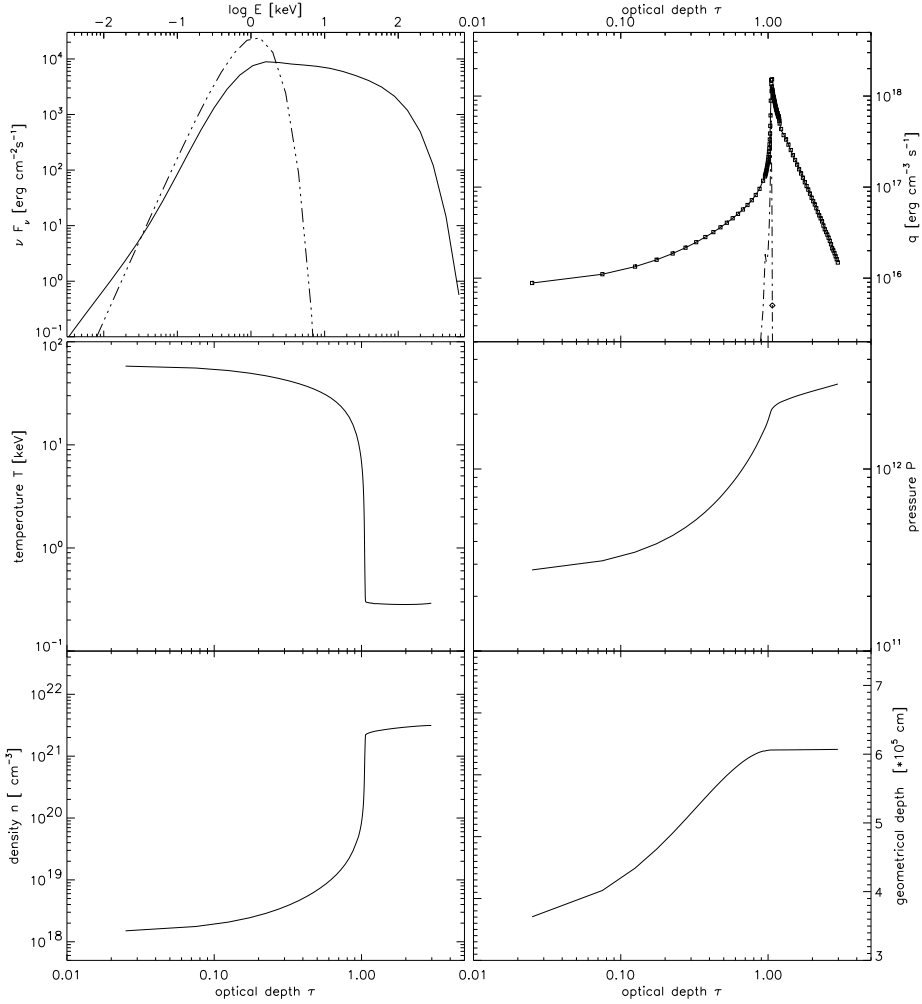} \caption{
   X-ray spectra of proton- illuminated optically thick cool disks. Upper
    left panel: emergent model spectrum (solid line) and blackbody
    with temperature according to proton energy flux (dotted--dashed
    line); upper right panel: combined heating rates from proton
    heating and electron thermal conductivity (solid line) and
    radiative cooling rates due to Comptonization and bremsstrahlung
    (squares) - the dotted line shows the rates from electron
    conductivity alone; lower panels show from left to right and from
    top to bottom the electron temperature $T_\mm{e}$, pressure P,
    electron density $n_\mm{e}$ and the geometrical depth of the layer
    $z/R$, for a solution with $F_\mm{p} = 8.1\times 10^{21}$ erg
    cm$^{-2}$ s$^{-1}$, $T_\mm{p} = 15$ MeV, $M = 8 M_{\sun}$, $R = 10
    R_\mm{g}$ } \label{fig:coolall}
\end{figure*}

For the thick disks we use the same model parameters as in paper I,
i.e. the accretion luminosity $L=0.1L_\mm{Edd}$, the viscosity
parameter $\alpha=0.1$ and the fraction of the energy released in the
virialized atmosphere $f=0.95$. The mass of the galactic BH is
$M_\mm{BH}=8M_\odot$ and the mass of the AGN is
$M_\mm{AGN}=8\times10^6M_\odot$.

Fig. \ref{fig:coolall} shows the solution at $r=10$ for a proton energy
flux of $F_\mm{p} = 8.1\times 10^{21}$ erg cm$^{-2}$ s$^{-1}$,
$T_\mm{p} = 15.6$ MeV. The cool disk is separated by a sharp
temperature front from a hot part at an optical depth of order
unity. This step like temperature profile is much more pronounced
compared to the results of paper I.  The temperature of the hot part
is $T_\mm{e}\approx60$~keV. The hot layer acts as an effective
Comptonizing region and a hard spectrum is emitted. The power law
index of the BH spectra in $E F(E)$ is $s\approx 0.2$. At the soft end
of the spectrum there is an excess of soft photons with respect to a
blackbody with equivalent energy flux. This excess is due to a
`reverse photosphere effect', as explained in \cite{deufel01}.

The upper right panel of Fig. \ref{fig:coolall} shows the heating and
cooling rates. In the narrow transition zone ($\Delta\tau\approx 0.1$
) between the hot top and the cool bottom electron conductivity is not
negligible. Elsewhere conductivity plays no role for the energy
balance. 

The geometrical depth of the heated layer is
$z_\mm{tot}\approx6\times10^5$~cm. This is small compared to the
distance from the compact object ($R=2.3\times10^7$~cm). Thus a cool
disk with a proton heated skin on top of it can still be considered a
``thin' disk.

\begin{figure*}[t]
\mbox{
\includegraphics[keepaspectratio=false,width=\columnwidth]{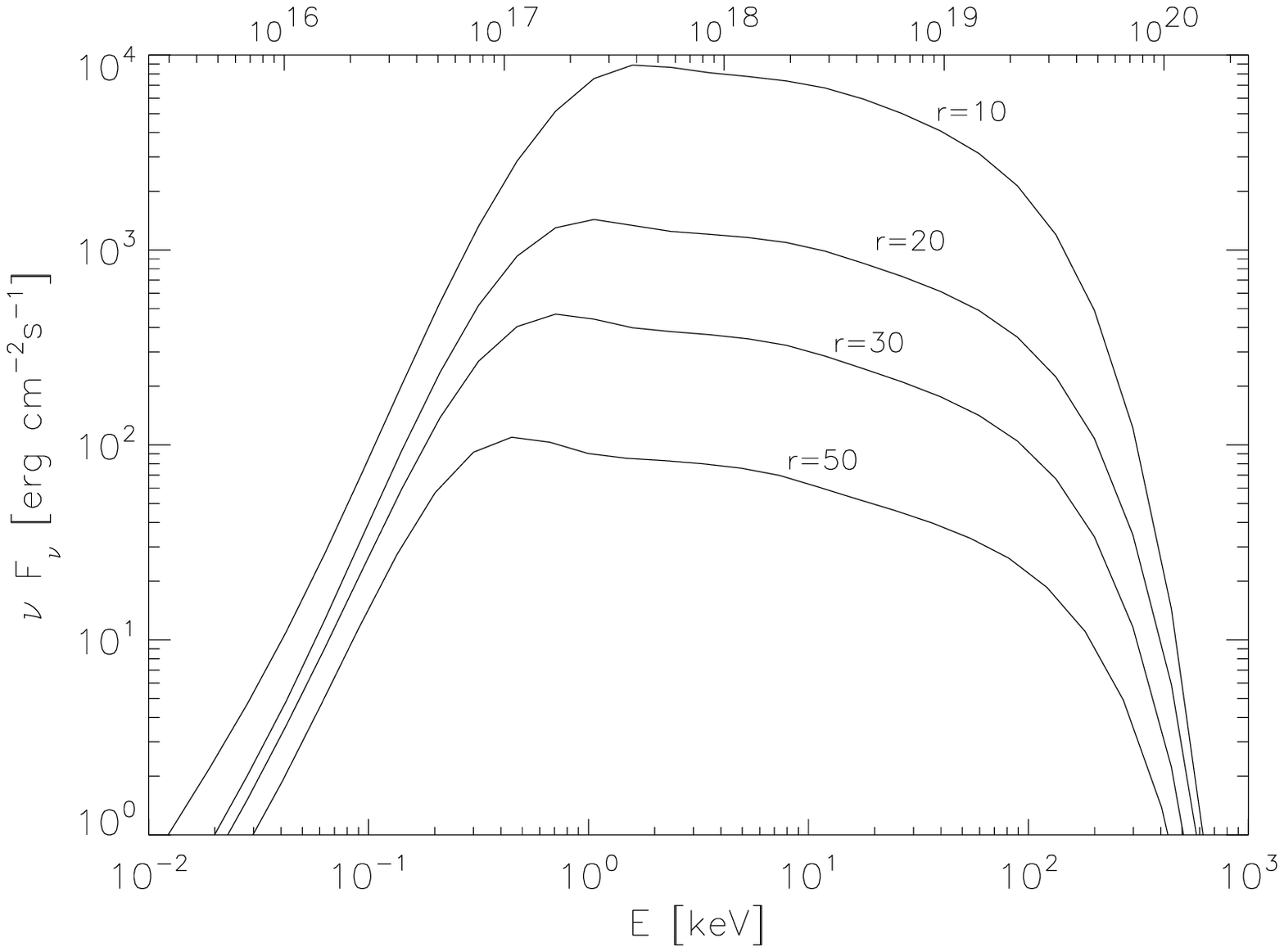}
\includegraphics[keepaspectratio=false,width=\columnwidth]{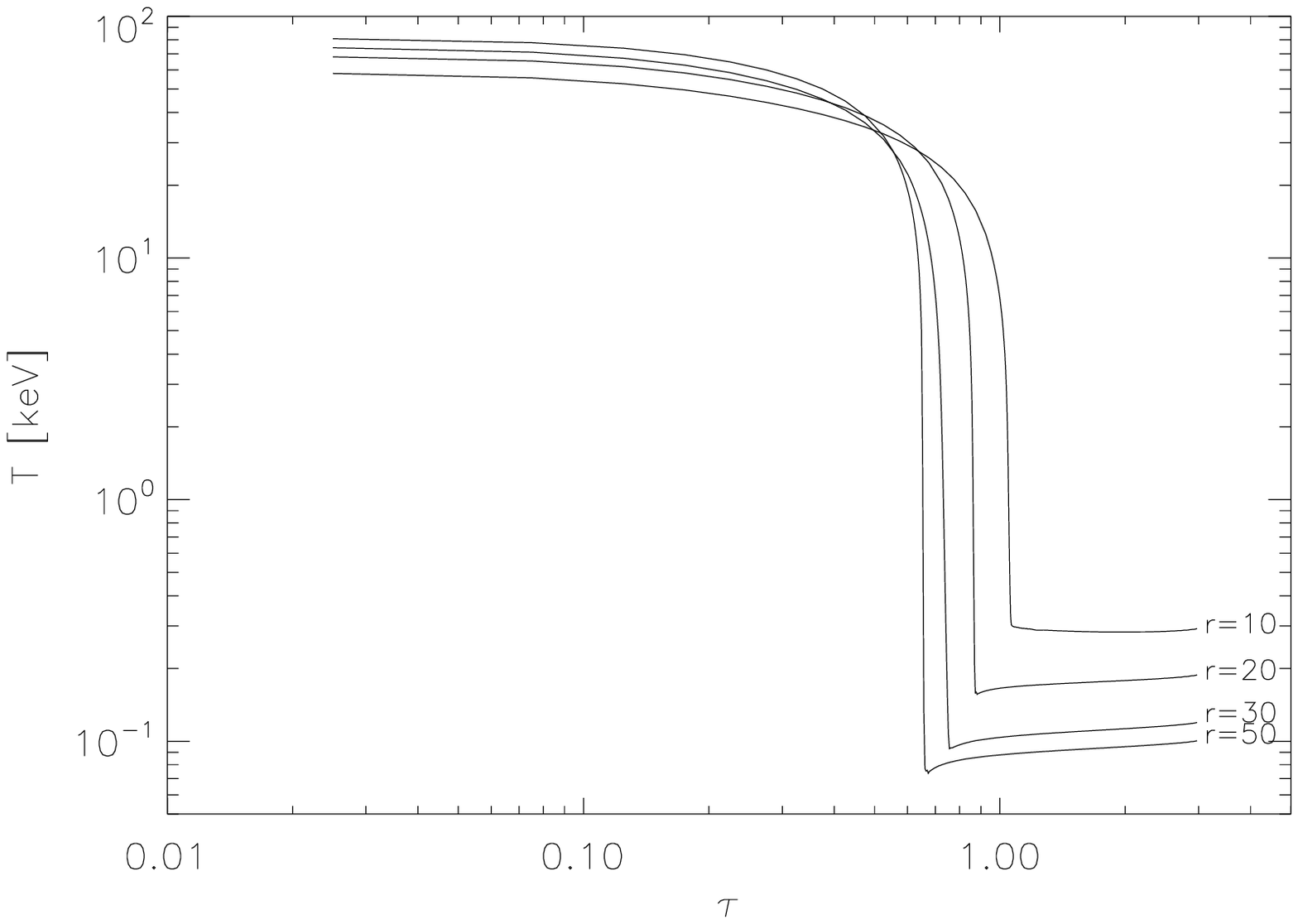}
}  \caption{Dependence of the emergent model spectra and the
temperature profiles on the distance from a galactic BH with
$M=8M_\odot$. The surface temperature of the of the disk increases slightly with distance, whereas the optical depth of the hot surface layer decreases.}
\label{bhsol}
\end{figure*}

Fig. \ref{bhsol} shows the dependence of the emergent spectrum and the
temperature profile on the distance from the compact object in the
galactic BH case. We obtain a result comparable to paper I, i.e. with
increasing distance the temperature of the hot part increases
somewhat, whereas the optical depth of the heated layer decreases. A
discussion of these effects can be found in paper I.

For distances $R>50 R_\mm{S}$ from the galactic BH we could not find
heated solutions. With increasing distance from the central object the
proton temperature ($\propto R^{-1}$) as well as the proton energy
flux ($\propto R^{-3}$) decrease. In these conditions we find
solutions in which the energy supplied by the incident protons is
thermalized directly in the cool disk by bremsstrahlung, without
forming a hot surface layer.  It is likely that a very thin ($\tau \ll 0.01$) hot 
layer still forms in these conditions, but we are unable to resolve it with the
present method. This is indicated by the existence of thin hot atmosphere
solutions at low incident energy flux in the case of protons heating a
neutron star surface \citep{zampi95,deufel01}. It probably also occurs
in the present case, but since these thin layers at large distances
from the hole do not contribute much to the overall spectrum, we have
not pursued this further.

\begin{figure*}[t]
  \mbox{
    \includegraphics[keepaspectratio=false,width=\columnwidth]{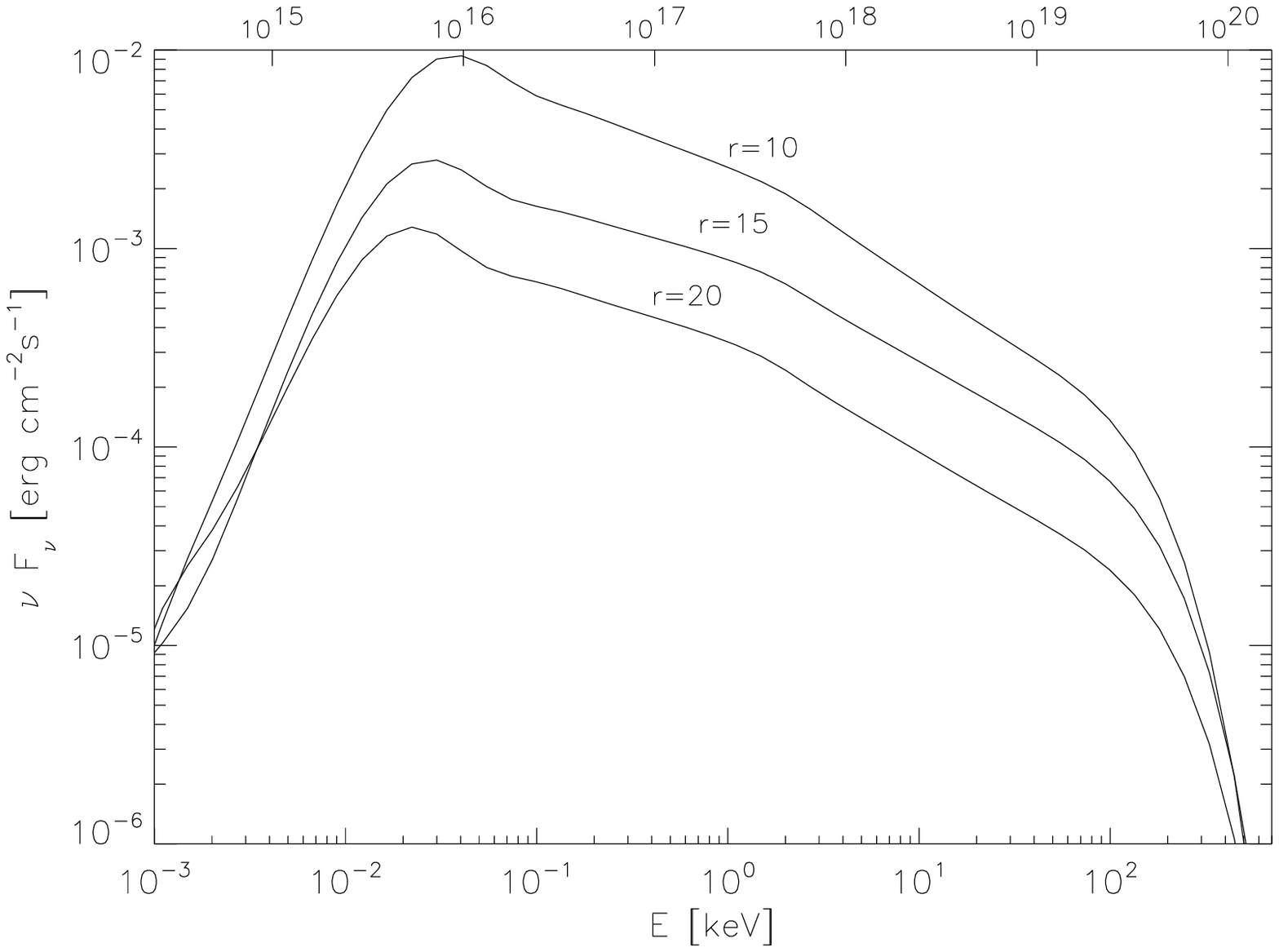}
    \includegraphics[keepaspectratio=false,width=\columnwidth]{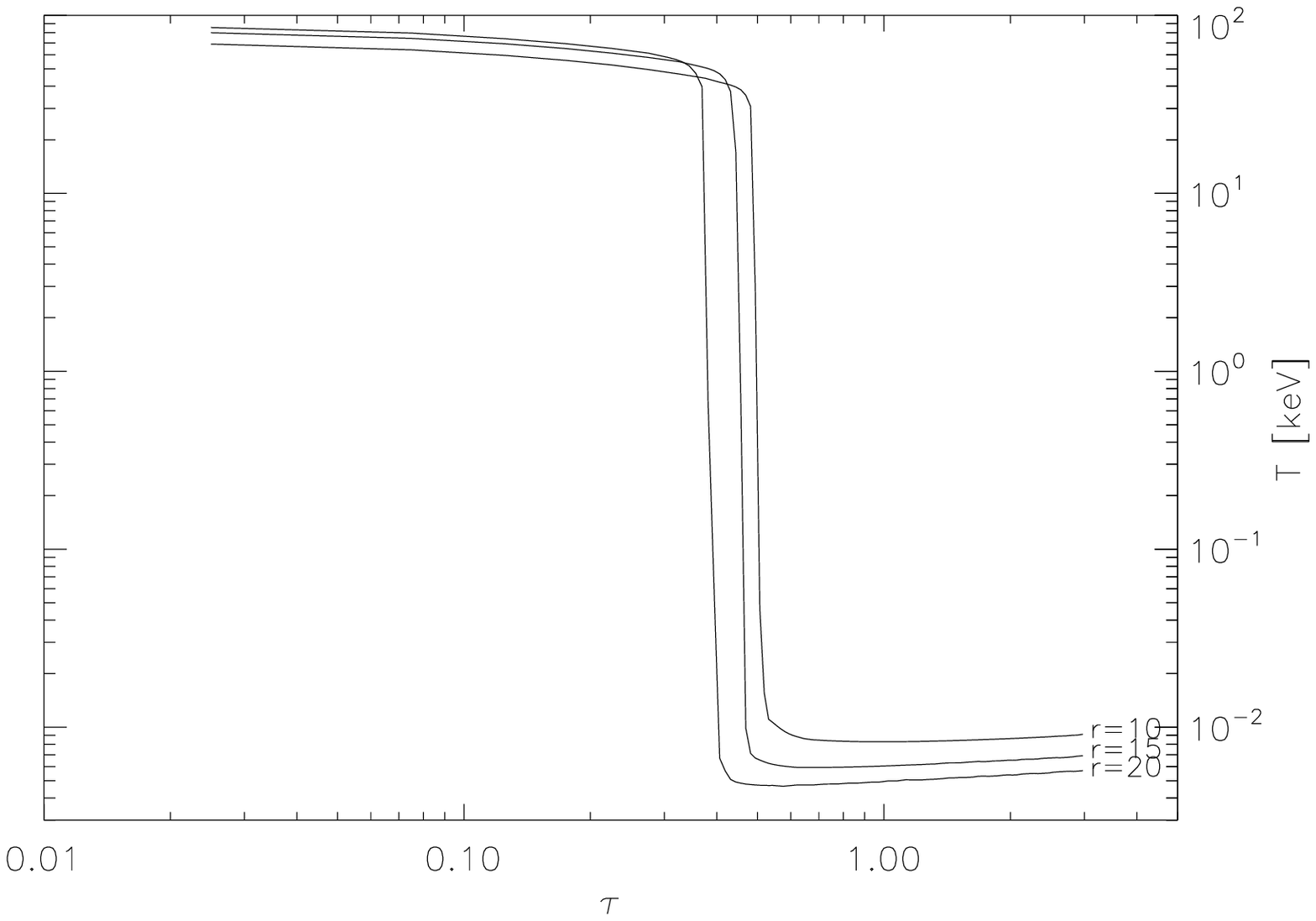}
    } \caption{Dependence of the emergent model spectra and the
    temperature profiles on the distance from the hole in the AGN case
    with $M=8\times10^6 M_\odot$. The optical depth of the hot part is
    much smaller than in the galactic BH cases.}
\label{agnsol}
\end{figure*}

Fig. \ref{agnsol} shows the dependence of the emergent spectrum and
the temperature profile on the distance from the compact object in the
AGN case.  We obtain slightly different solutions compared to our
results in paper I.  The optical depth of the heated layer is
smaller ($\tau\lesssim 0.5$), the temperature of the hot part is
$T_\mm{e}\approx$ 70--90 keV. Again a steep temperature jump separates
it from the cool disk underneath.  Due to the smaller optical depth of
the heated layer Comptonization is not as efficient. A significant fraction
of the incident ions passes through the hot layer and thermalizes in the
cool disk underneath. This increases the soft photon flux relative to
the Comptonized flux, and produces a steeper spectrum. The
index of the AGN spectra in $E F(E)$ is $s\approx 0.6$. Our model does
not produce heated solutions at distances $R>20 R_\mm{S}$ from the compact
object, though heated atmospheres might exist at very small optical
depths (see note above).

\section{Proton illumination of disks with moderate optical depth}
\label{sec:slim}

\begin{figure*}
  \sidecaption \includegraphics[width=12cm]{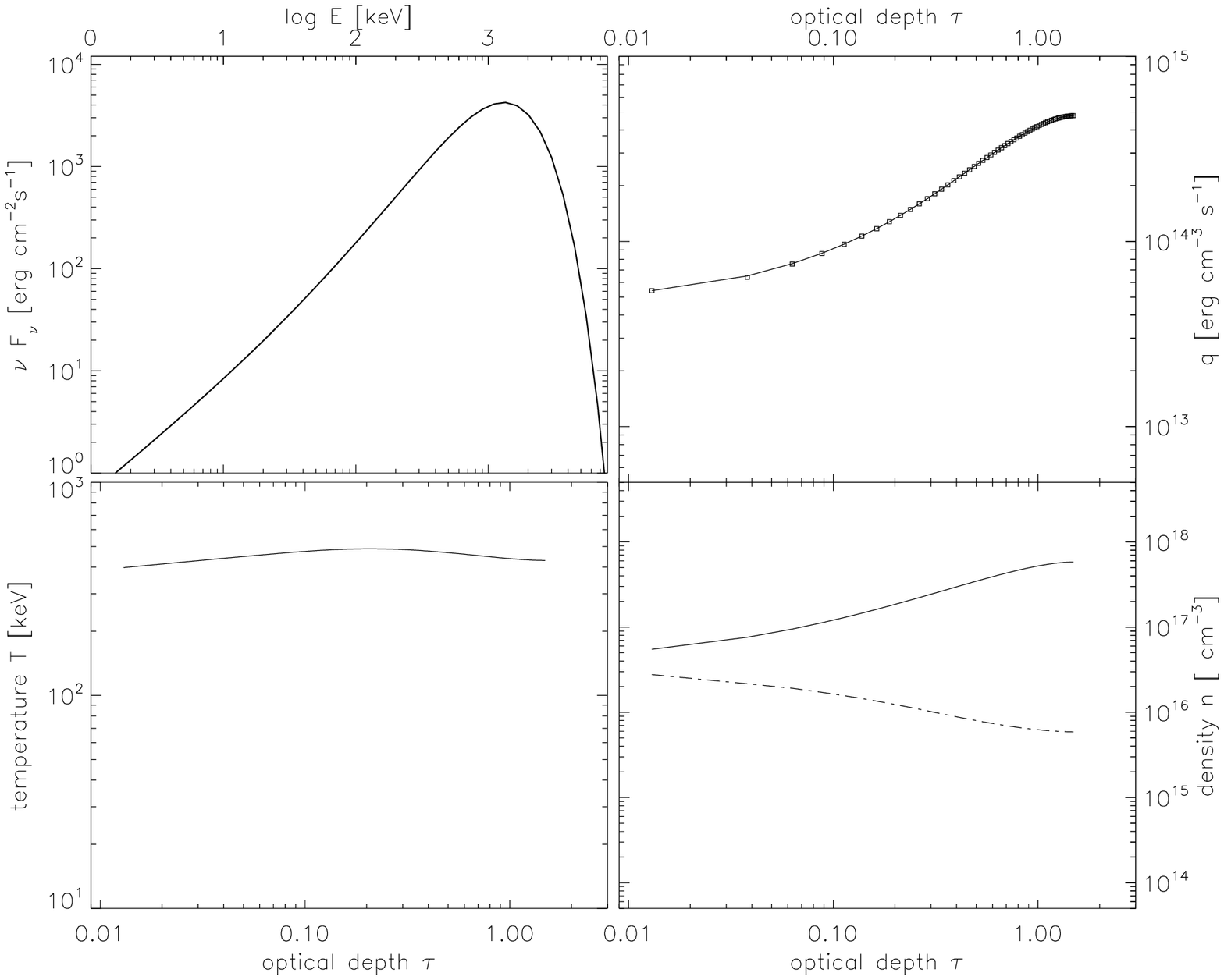} \caption{
A `warm disk', of optical depth 1.5, externally heated by ion illumination. 
  Upper left panel: emergent model spectrum; upper right panel: heating
  rates from proton heating (solid line) and combined radiative
  cooling rates from Comptonization, bremsstrahlung and pair processes
  (squares); lower left panel: electron temperature $T_\mm{e}$; lower
  right panel: proton number density (solid line) of the layer and
  pair number density (dashed line).} \label{slimall}
\end{figure*}

\begin{figure}[th]
\includegraphics[width=\hsize]{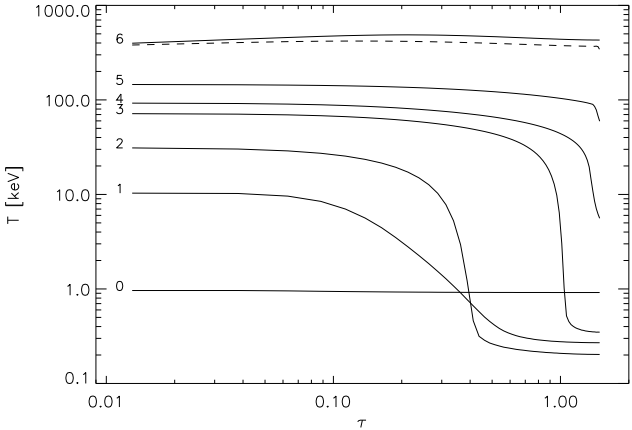}
\caption{Evolution from a cool disk to a warm disk by exposure to 
  virialized protons at $r=10$.  Initial temperature $T_\mm{e}=1$keV
  The numbers at the lines denote the sequential stages, referred to
  in the text. The final stage [6] shows an equilibrium at $T_\mm{e}
  \approx 400$ keV. The dashed line shows the temperature profile
  after one dynamical time-scale.}
\label{tevol}
\end{figure}

\begin{figure}
\includegraphics[width=\hsize]{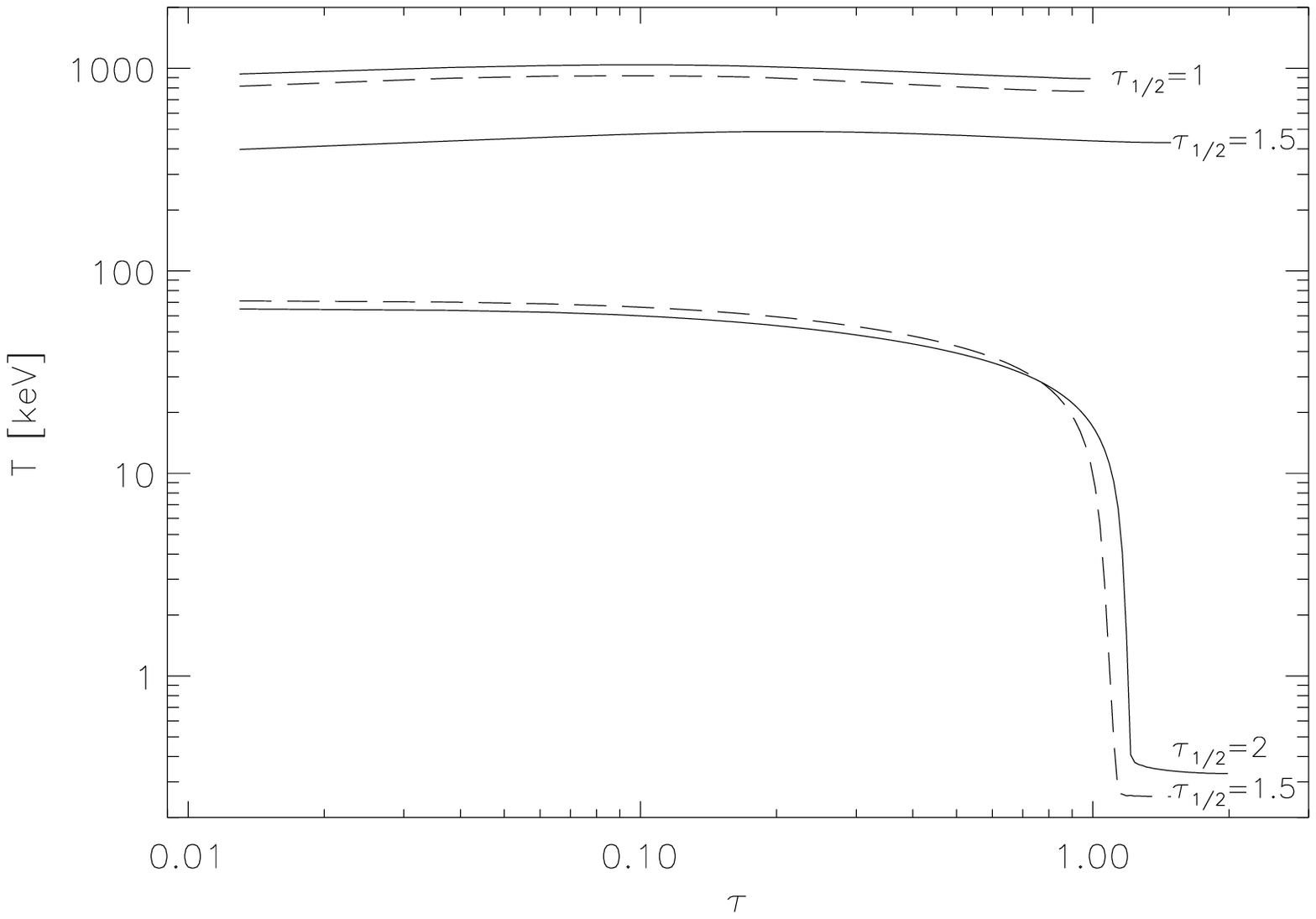}
\caption{Equilibrium temperature profiles of warm disks
exposed to virialized protons at $r=10$ (solid line) and
  $r=15$ (dashed line) with different initial optical depths
  $\tau_{1/2}=2,1.5,1.$ for $r=10$ and $\tau_{1/2}=1.5,1$ for $r=15$.
  }
\label{diffdep}
\end{figure}

The temporal evolution of an initially cool ($T_\mm{e}=1$~keV), thin
disk with an optical depth $\tau_{1/2}=1.5$ (measured from the surface
to the mid-plane) at $r=10$  for the galactic 
black hole candidate (BHC) case is shown in
Fig. \ref{tevol}. The energy flux from the virialized protons is
$F_\mm{p} = 8.1\times10^{21}$ erg cm$^{-2}$ s$^{-1}$. In the beginning
the temperature of the top layers increases due to the impinging hot
protons, whereas the mid-plane region cools due to bremsstrahlung
[stage 0--2, Fig.\ref{tevol}].  As the top layers are heated, the
stopping power of the plasma decreases and the protons penetrate
deeper into the disk.  Eventually hot protons reach the mid-plane
region and proton heating overcomes bremsstrahlung cooling even there
[3]. At this stage no effective cooling mechanism is present and the
temperature continues to rise everywhere in the disk [4-5]. At
temperatures $kT \gtrsim 200$~keV pair production becomes more and
more important. The extra electrons serve as additional scattering
partners for the Coulomb collisions with the protons and the photons
of the radiation field. Thus pair production limits the maximum
attainable temperatures, and the disk adjusts to a new equilibrium
state [6] at a temperatures $T_\mm{e}\simeq 400$ keV.  The dashed line
in Fig.  \ref{tevol} shows the temperature profile after one dynamical
time-scale at that radius, $t_\mm{d}=1/\Omega_\mm{K}\simeq
3.5\times10^{-3}$s.

Fig. \ref{slimall} shows an overview of that solution. The pair
number density $z$ at the top of slab reaches $z\approx50$\% of the
proton number density and drops off at the mid-plane to $z\approx 1$\%.
The spectrum of such a thin proton heated disk peaks at $\approx
1000$~keV. We refer to those disks as \em warm disks \rm. They are
still considerably cooler than the local virial temperature.

Fig. \ref{diffdep} shows the dependence of the equilibrium temperature
profile on the initial optical depth of the layer and the distance
from the central object.  With increasing distance the proton energy
flux as well as the proton penetration depth decreases. A cool disk
can therefore be transferred into the warm state only within a certain
distance from the BH. If the hot protons do not reach the mid-plane
anymore, a cool interior can be maintained which looses its energy
very efficiently via bremsstrahlung, as the hot layers above are
optically thin.  Our model computations show that at $r=15$ a thin
disk can maintain a cool interior for $\tau_{1/2}=1.5$ whereas at
$r=10$ a disk with $\tau_{1/2}=1.5$ switches into the warm state.

The temperature of the warm state also depends on both the distance
and the overall optical depth of the layer. For $\tau_{1/2}=1$ and
$r=10$ our model predicts a temperature of $\approx 1$~MeV. At such
temperatures our classical proton heating formalism starts deviating
from the correct relativistic expression. The classical treatment
underestimates the proton--electron heating rates at high temperatures
\citep{deufel01}. But further pair processes and radiative cooling
terms should also be included (see below), which again limits the
maximum temperatures.

The transition from a cool disk to the warm state also takes place in
AGNs. For the above AGN parameters (see Sect. \ref{sec:protill}) we
find the transition to occur for $\tau_{1/2} \lesssim 0.2$ at $r
\lesssim 10$.  The temperatures of the warm state are in the MeV
range.  At such temperatures ($\theta>1$) our treatment of the pair
processes and the radiation field needs more scrutiny. Further pair
production processes should be included ($\gamma e \rightarrow e
e^{+}e^{-}$, $e p \rightarrow e p e^{+}e^{-}$) as well as additional
radiative cooling terms (bremsstrahlung from $e^{+}e^{-}$ and $e^{\pm}
e^{\pm}$ collisions). These will further limit the maximum attainable
temperatures.  Therefore we do not think that temperatures of several
MeV found in our simulation of the warm state in AGNs are realistic.  The
important result of our investigation is that the transition is not
only restricted to the BHCs but also takes place in AGNs. But compared
to the BHCs the transition in the AGNs occurs in a more narrow zone
around the BH (in terms of Schwarzschild radii) and the vertical
extend (in terms of Thomson optical depths) is smaller.

\section{Discussion and conclusions}

We have investigated a model for the production of a hard spectral
component ($h\nu\gg 1$keV) from a cool disk around a compact object.
The model assumes that a standard cool accretion disk is surrounded by
a hot atmosphere. This atmosphere can either be due to an ion
supported accretion flow (ADAF) which partially overlaps in radial
extent with the cool disk or to a hot corona which is e.g.
magnetically heated \citep[e.g.][]{matteo99}.  For the actual
calculations presented here, we have used parameters representative
for ADAFs. Fast protons from this atmosphere interact with the cool
disk material and produce a warm ($\sim 80$ keV) surface layer, where
soft photons from the cool disk are Comptonized. We have considered
two cases in detail: a disk around an $8 M_\odot$ black hole (galactic
BHC case), and an AGN case with an $8\cdot10^6 M_\odot$ black hole.

We find that the response of the cool disk to proton illumination
depends on its optical thickness. At sufficient thickness
($\tau\gtrsim 1.5$) the cool disk is able to effectively reprocess
incoming energy flux into soft photons. Here, a two-layer structure
develops, consisting of a warm surface layer of depth $\tau\sim 1$ on
top of a cool disk. At lower total optical depth, the soft photon
emissivity is insufficient to cool the disk and the disk heats up, by
the proton energy flux, to an approximately isothermal layer of
intermediate (a few 100 keV) temperatures. We call these cases {\it
  warm disks}: their temperature is higher than an accretion disk
radiating as a black-body, but still much lower than the virial
temperature.

In the optically thick case, the energy balance between the warm
surface layer and the cool disk is like that in the Haardt-Maraschi
(1991, 1993) model of a corona over a cool disk. For geometrical
reasons about half of the Comptonized photons is reprocessed by the
cool disk, the other escapes. The Compton $y-$parameter adjusts such
that the amplification factor matches this requirement. This balance
regulates the temperature of the layer, while its optical depth is
determined by the penetration depth of the protons. The balance is
strongly self-regulating: the temperature and optical depth of the
warm layer depend little (within limits) on temperature and energy
flux of the incident protons, so that also the shape of the emergent
X-ray spectrum varies little (see Haardt and Maraschi 1993, and paper
I for a more detailed discussion).

We find that the existence of these proton-heated surface layers is
limited to the innermost regions $r\lesssim 50$ of the disk. At
larger distances from the hole, the optical depth of the warm layer
drops rapidly, and its contribution to the spectrum becomes small.

For our BHC case the temperature and optical depth of the warm layer
are in the right range to produce spectra with the main features of
black hole binaries in their hard states. In the AGN case, the
Comptonization is weaker, and the hard spectral component
somewhat steeper (photon index around 2.6) than in the BHC case (index
around 2.2). In both the BHC and AGN cases the spectra cut off around
200 keV. The maximum distance from the hole where a significant warm
layer is formed is somewhat smaller in the AGN case ($R_{\rm
max}/R_{\rm S}\approx 20$) than in the BHC case ($R_{\rm max}/R_{\rm
S}\approx 50$). This is due to the lower energy flux per unit of
surface area, and the reduced penetration depths in the cooler AGN
disks (see also Deufel et al. 2001).

Near the inner edge of a cool accretion disk (at the distance $R_{\rm
  i}$ from the hole), its optical depth becomes small (in the thin
disk limit, it varies by the well known factor $f=[1-(R_{\rm
  i}/R)^{1/2}$]\,). Though the optical depth of a cool accretion disk
is generally substantial at the accretion rates inferred for BHC and
AGN, there is always a region of low optical depth near the inner
edge. This is also where the disk would be exposed to the largest
proton energy flux from an ADAF or corona. At optical depths
$\tau_{1/2}\lesssim 1.5$ we find that an initially cool disk heats up
to a new equilibrium with nearly uniform temperatures of a few 100 keV
(Fig. 5). The transition takes place in a time of the order of the
dynamical time scale $1/\Omega$. Initially, the protons are stopped in
the cool layers before reaching the midplane. Due to the thermal
expansion of the plasma the electron densities drop, and as a
consequence the proton heating finally overcomes the cooling by
bremsstrahlung ($\sim n_{\rm e}^2$). A new equilibrium is found when
pair production sets in. The pairs add to the optical depth of the
layer and increase the radiative loss of the plasma (by
Comptonization) until it matches the input by proton heating.

Our solutions are local equilibria for conditions at a given distance
from the hole. In disks in which the optical depth is low ($\lesssim 1$)
near the inner edge but at least a few at larger distance,
proton heating would produce our warm disk structure of approximately
uniform temperature in a (possibly narrow) zone near the inner edge.
Further out it would instead produce a warm surface layer overlying
the cool disk. Both regions would contribute to the overall X-ray
spectrum.

Since the energy flux by proton heating is highest near the inner
edge, the warm disk component at that location could contribute
significantly to the overall spectrum even if its radial extent is
limited.  In our solutions, it would add a very flat component (photon
index around 1, up to about 1MeV, Fig. 4), so the overall spectral
index could be significantly less than 2, as observed in some of the
hardest spectra. The warm surface layers over the cool disk do not
have such flat spectra (Fig.  2), and alone can not explain the
hardest observed spectra (with indices around 1.5).  The combination
has the potential to explain the entire range of observed hard
spectra, but for a quantitative result the radial structure of a
proton-illuminated disk needs to be treated in more detail.

In this context we note that our warm disk solutions ignore a
potentially important cooling mechanism, namely Compton cooling by
soft photons from a cool disk. Though there are no cool regions
internally in the warm disk, the nearby cool disk at larger radii
produces a soft photon flux. A fraction of this flux could travel
radially, illuminating the warm disk region. The flux of such soft
photons is probably small, since the disk is thin at all radii and the
radial optical depths therefore large. For quantitative assessment a
two-dimensional radiative transfer model has to be developed, which is
outside the scope of the present treatment.

The warm disk solution has a combination of temperature and optical
depth just in the region where a normal, internally heated disk cannot
exist. In such disks, the ($T,\tau$) combination characterizing our
warm disks would lie on the SLE (Shapiro-Lightman-Eardley) branch of
accretion flows, which is thermal/viscously unstable. The combination
is quite stable in our warm disk solution precisely because it is not
heated internally but externally by the incident protons. In this
context it is interesting that model fits of low/hard states in terms
of stationary ADAF models sometimes yield parameter combinations on
the unstable SLE branch (for example in XTE 1118 +480 as analyzed by
Frontera et al. 2001). We interpret this as a strong
indication for the existence of our warm disk solutions.

The physics of the warm disk solutions presented here has further
interesting consequences. In anticipation of results to be given in
more detail in a future paper, we note here that at the temperatures
and densities in our warm disks, the time scale for establishing
thermal equilibrium between the electrons and the protons in the disk
(not the illuminating protons) is not short compared to the dynamical
time scale any more. Any mechanism which now provides energy to the
disk protons faster than they can exchange energy with the disk
electrons would give rise to a heating of the disk protons and,
depending on the temperature and density dependence of such a process,
might lead to runaway heating of the protons. Two energy channels
immediately come to mind: proton-proton interactions, i.e. heating of
the disk protons by the penetrating external protons, and internal
viscous heating according to standard disk theory.  We have neglected
these here, as their energetic contribution is negligible in our model
(the main energy channel is from the incident protons to the radiation
field, via the disk electrons).  If such an instability exists, it
might heat the disk protons sufficiently for the disk to expand and
feed its mass into the surrounding ADAF.

\begin{acknowledgements}
  This work was done in the research network `Accretion onto black
  holes, compact stars and proto stars' funded by the European
  Commission under contract number ERBFMRX-CT98-0195.
\end{acknowledgements}

\bibliographystyle{apj}
\bibliography{aamnem99,pub}

\begin{thebibliography}{34}
\expandafter\ifx\csname natexlab\endcsname\relax\def\natexlab#1{#1}\fi

\bibitem[{{Aharonian} \& {Sunyaev}(1984)}]{aha}
{Aharonian}, F. \& {Sunyaev}, R. 1984, MNRAS, 210, 257

\bibitem[{{Alme} \& {Wilson}(1973)}]{alme}
{Alme}, M.~L. \& {Wilson}, J.~R. 1973, ApJ, 186, 1015

\bibitem[{{Chandrasekhar}(1942)}]{chandra}
{Chandrasekhar}, S. 1942, "Principles of Stellar Dynamics" (University of
  Chicago Press, 1992.), Ch.2 and Sect. 5.6

\bibitem[{{Deufel} {et~al.}(2001){Deufel}, {Dullemond}, \& {Spruit}}]{deufel01}
{Deufel}, B., {Dullemond}, C.~P., \& {Spruit}, H.~C. 2001, A\&A, 377, 955

\bibitem[{{Deufel} \& {Spruit}(2000)}]{deufel00}
{Deufel}, B. \& {Spruit}, H.~C. 2000, A\&A, 362, 1, paper I

\bibitem[{{di Matteo} {et~al.}(1999){di Matteo}, {Celotti}, \&
  {Fabian}}]{matteo99}
{di Matteo}, T., {Celotti}, A., \& {Fabian}, A.~C. 1999, MNRAS, 304, 809

\bibitem[{{Dove} {et~al.}(1997){Dove}, {Wilms}, {Maisack}, \&
  {Begelman}}]{dove97}
{Dove}, J.~B., {Wilms}, J., {Maisack}, M., \& {Begelman}, M.~C. 1997, ApJ, 487,
  759+

\bibitem[{{Esin} {et~al.}(1997){Esin}, {McClintock}, \& {Narayan}}]{esin97}
{Esin}, A.~A., {McClintock}, J.~E., \& {Narayan}, R. 1997, ApJ, 489, 865+

\bibitem[{{Haardt} \& {Maraschi}(1991)}]{haardt91}
{Haardt}, F. \& {Maraschi}, L. 1991, ApJ Lett., 380, L51

\bibitem[{{Haardt} \& {Maraschi}(1993)}]{haardt93}
---. 1993, ApJ, 413, 507

\bibitem[{{Haardt} {et~al.}(1994){Haardt}, {Maraschi}, \& {Ghisellini}}]{hmg93}
{Haardt}, F., {Maraschi}, L., \& {Ghisellini}, G. 1994, ApJ, 432, L95

\bibitem[{{Haardt} {et~al.}(1997){Haardt}, {Maraschi}, \& {Ghisellini}}]{hmg97}
---. 1997, ApJ, 476, 620

\bibitem[{{Jauch} \& {Rohrlich}(1976)}]{jauch76}
{Jauch}, J.~M. \& {Rohrlich}, F. 1976, The theory of photons and electrons. The
  relativistic quantum field theory of charged particles with spin one-half
  (Texts and Monographs in Physics, New York: Springer, 1976, 2nd ed.)

\bibitem[{{Mart\'{\i}n} {et~al.}(1994{\natexlab{a}}){Mart\'{\i}n}, {Rebolo},
  {Casares}, \& {Charles}}]{martin94a}
{Mart\'{\i}n}, E., {Rebolo}, R., {Casares}, J., \& {Charles}, P.
  1994{\natexlab{a}}, ApJ, 435, 262

\bibitem[{{Mart\'{\i}n} {et~al.}(1994{\natexlab{b}}){Mart\'{\i}n}, {Spruit}, \&
  {van Paradijs}}]{martin94b}
{Mart\'{\i}n}, E., {Spruit}, H., \& {van Paradijs}, J. 1994{\natexlab{b}},
  A\&A, 291, 43

\bibitem[{{Nakamura} \& {Osaki}(1993)}]{nakamura93}
{Nakamura}, K. \& {Osaki}, Y. 1993, PASJ, 45, 775

\bibitem[{{Narayan} \& {Yi}(1994)}]{narayan94}
{Narayan}, R. \& {Yi}, I. 1994, ApJ Lett., 428, L13

\bibitem[{{Narayan} \& {Yi}(1995{\natexlab{a}})}]{narayan95a}
---. 1995{\natexlab{a}}, ApJ, 444, 231

\bibitem[{{Narayan} \& {Yi}(1995{\natexlab{b}})}]{narayan95b}
---. 1995{\natexlab{b}}, ApJ, 452, 710

\bibitem[{{Pozdnyakov} {et~al.}(1983){Pozdnyakov}, {Sobel}, \&
  {Syunyaev}}]{pozd83}
{Pozdnyakov}, L.~A., {Sobel}, I.~M., \& {Syunyaev}, R.~A. 1983, Astrophysics
  and Space Physics Research, 2, 189

\bibitem[{{Rees} {et~al.}(1982){Rees}, {Phinney}, {Begelman}, \&
  {Blandford}}]{rees82}
{Rees}, M.~J., {Phinney}, E.~S., {Begelman}, M.~C., \& {Blandford}, R.~D. 1982,
  Nat, 295, 17

\bibitem[{{Ross} \& {Fabian}(1993)}]{ross93}
{Ross}, R.~R. \& {Fabian}, A.~C. 1993, MNRAS, 261, 74

\bibitem[{{Shakura} \& {Sunyaev}(1973)}]{sunyaev73}
{Shakura}, N.~I. \& {Sunyaev}, R.~A. 1973, A\&A, 24, 337

\bibitem[{{Shapiro} {et~al.}(1976){Shapiro}, {Lightman}, \&
  {Eardley}}]{shapiro76}
{Shapiro}, S.~L., {Lightman}, A.~P., \& {Eardley}, D.~M. 1976, ApJ, 204, 187

\bibitem[{{Spruit}(1997)}]{spruit97}
{Spruit}, H.~C. 1997, in Berlin Springer Verlag Lecture Notes in Physics, Vol.
  487, 67--76

\bibitem[{{Spruit} \& {Haardt}(2000)}]{spruit00}
{Spruit}, H.~C. \& {Haardt}, F. 2000, MNRAS, 315, 751

\bibitem[{{Stepney} \& {Guilbert}(1983)}]{stepg83}
{Stepney}, S. \& {Guilbert}, P.~W. 1983, MNRAS, 204, 1269

\bibitem[{{Sunyaev} \& {Titarchuk}(1980)}]{titar80}
{Sunyaev}, R.~A. \& {Titarchuk}, L.~G. 1980, A\&A, 86, 121

\bibitem[{{Svensson}(1982)}]{svensson82}
{Svensson}, R. 1982, ApJ, 258, 335+

\bibitem[{{Svensson} \& {Zdziarski}(1994)}]{svensson94}
{Svensson}, R. \& {Zdziarski}, A.~A. 1994, ApJ, 436, 599

\bibitem[{{Turolla} {et~al.}(1994){Turolla}, {Zampieri}, {Colpi}, \&
  {Treves}}]{turolla94}
{Turolla}, R., {Zampieri}, L., {Colpi}, M., \& {Treves}, A. 1994, ApJ Lett.,
  426, L35

\bibitem[{{Zampieri} {et~al.}(1995){Zampieri}, {Turolla}, {Zane}, \&
  {Treves}}]{zampi95}
{Zampieri}, L., {Turolla}, R., {Zane}, S., \& {Treves}, A. 1995, ApJ, 439, 849

\bibitem[{{Zane} {et~al.}(1995){Zane}, {Turolla}, {Zampieri}, {Colpi}, \&
  {Treves}}]{zane95}
{Zane}, S., {Turolla}, R., {Zampieri}, L., {Colpi}, M., \& {Treves}, A. 1995,
  ApJ, 451, 739+

\bibitem[{{Zel'dovich} \& {Shakura}(1969)}]{zel}
{Zel'dovich}, Y.~B. \& {Shakura}, N.~I. 1969, Soviet Astron.-AJ, 13, 175,
  {ZS69}

\end{thebibliography}

\end{document}